\documentstyle[12pt]{article}

\begin{document}

\begin{titlepage}

\begin{flushright}
SLAC--PUB--6263\\
June 1993\\
T/E
\end{flushright}

\vspace{0.5cm}

\begin{center}
\huge\bf Heavy Quark Symmetry
\end{center}

\vspace{0.5cm}

\begin{center}
Matthias Neubert\footnote{Supported by the Department of Energy under
contract DE-AC03-76SF00515.}\\
Stanford Linear Accelerator Center\\
Stanford University, Stanford, California 94309
\end{center}

\vspace{0.5cm}

\begin{abstract}
We review the current status of heavy quark symmetry and its
applications to weak decays of hadrons containing a single heavy quark.
After an introduction to the underlying physical ideas, we discuss in
detail the formalism of the heavy quark effective theory, including a
comprehensive treatment of symmetry breaking corrections. We then
illustrate some nonperturbative approaches, which aim at a dynamical,
QCD-based calculation of the universal form factors of the effective
theory. The main focus is on results obtained using QCD sum rules.
Finally, we perform an essentially model-independent analysis of
semileptonic $B$ meson decays in the context of the heavy quark
effective theory.
\end{abstract}

\bigskip
\centerline{(to appear in Physics Reports)}

\end{titlepage}

{\huge\bf Preface}\linebreak
\addcontentsline{toc}{chapter}{Preface}

A scan through the {\sl Review of Particle Properties\/} gives an
impression of the great variety of phenomena caused by the weak
interactions. Attempts to understand this rich phenomenology have led
to much progress in particle physics. Today, the standard model of the
strong and electroweak interactions provides a most successful
description of the physics currently accessible with particle
accelerators. In spite of its success, however, many open questions
remain. In particular, a large number of undetermined parameters are
associated with the flavor sector of the theory. These parameters are
the quark and lepton masses, as well as the four angles of the
Cabibbo-Kobayashi-Maskawa matrix, which describes the mixing of the
mass eigenstates of the quarks under the weak interactions. Quite
obviously, weak decays offer the most direct way to determine these
mixing angles and to test the flavor sector of the standard model. But
they are also ideally suited for a study of that part of strong
interaction physics which is least understood: the nonperturbative
long-distance forces, which are responsible for the confinement of
quarks and gluons into hadrons. Indeed, these two aspects cannot be
separated from each other. An understanding of the connection between
quark and hadron properties is a prerequisite for a precise
determination of the parameters of the standard model.

In many instances this puts important limitations on the amount of
information that can be deduced from weak interaction experiments at
low energies. The theoretical description of hadron properties often
relies on very naive bound state models with no direct connection to
the underlying theory of quantum chromodynamics (QCD). Indeed, at low
energies this theory has proved so intractable to analytical approaches
that reliable predictions can only be made based on symmetries. A
well-known example is chiral symmetry, which arises since the current
masses of the light quarks are small compared to the intrinsic mass
scale of the strong interactions given, say, by the mass of the proton.
It is then useful to consider, as a first approximation, the limiting
case of $n_f$ massless quarks, in which the theory has an ${\rm
SU}_L(n_f)\times{\rm SU}_R(n_f)$ chiral symmetry, which is
spontaneously broken to ${\rm SU}_V(n_f)$. Associated with this is a
set of massless goldstone bosons. This symmetry pattern persists in the
presence of small mass terms for the quarks. The goldstone bosons,
which also acquire small masses due to these perturbations, can be
identified with the light pseudoscalar mesons. The low energy theorems
of current algebra predict relations between the scattering and decay
amplitudes for processes involving a different number of these
particles, and chiral perturbation theory provides the modern framework
for a systematic analysis of the symmetry breaking corrections. This is
quite a powerful concept, which allows predictions from first
principles in an energy regime where the standard perturbative approach
to quantum field theory, i.e.\ an expansion in powers of the coupling
constant, breaks down. A nice illustration is provided by the
calculation of the matrix element of the flavor-changing vector current
between a kaon and a pion state. The Ademollo-Gatto theorem states
that, at zero momentum transfer, the corresponding form factor is
normalized up to corrections which are of second order in the symmetry
breaking parameter $m_s-m_u$. Chiral perturbation theory can be used to
calculate the leading corrections in an expansion in the light quark
masses in a model independent way. From such an analysis the form
factor can be predicted with an accuracy of 1\%. This makes it possible
to obtain a very precise determination of the element $V_{us}$ of the
Cabibbo-Kobayashi-Maskawa matrix from an analysis of the semileptonic
decay $K\to\pi\,\ell\,\bar\nu$.

Recently, it has become clear that a similar situation arises in the
opposite limit of very large quark masses. When the Compton wave length
$1/m_Q$ of a heavy quark bound inside a hadron is much smaller than a
typical hadronic distance of about 1 fm, the heavy quark mass is
unimportant for the low energy properties of the state. The strong
interactions of such a heavy quark with light quarks and gluons can be
described by an effective theory, which is invariant under changes of
the flavor and spin of the heavy quark. This ``heavy quark symmetry''
leads to similar predictions than chiral symmetry. In the limit
$1/m_Q\to 0$, relations between decay amplitudes for processes
involving different heavy quarks arise, and matrix elements of the
flavor-changing weak currents become normalized at the point of zero
velocity transfer. There is even an analog of the Ademollo-Gatto
theorem: In certain cases the leading symmetry breaking corrections are
of second order in $1/m_Q$. The existence of an exact symmetry limit of
the theory increases the prospects for a precise determination of the
element $V_{cb}$ of the quark mixing matrix. Ultimately, it can help to
promote the description of weak decays of hadrons containing a heavy
quark from the level of naive quark models to a theory of strong
interactions. The hope is to start from the model independent relations
that are valid in the limit of infinite heavy quark masses, and to
include the leading symmetry breaking corrections in an expansion in
$1/m_Q$. The theoretical framework for such an analysis is provided by
the so-called heavy quark effective theory.

In this review we present the current status of heavy quark symmetry
and of the heavy quark effective theory, with emphasis on the weak
decays of hadrons containing a single heavy quark. For these processes
the theory has already been developed to an extent which is no less
elaborate than, e.g., the modern formulation of chiral perturbation
theory. In particular, the symmetry breaking corrections have been
analyzed in great detail, and many quantitative predictions of
sometimes remarkable accuracy can be made. This review can be divided
into four parts: In the first two chapters we present the physical
picture and the ideas behind heavy quark symmetry in an intuitive,
introductory way. Similar introductions have been given by
Georgi,\footnote{in: Proceedings of TASI--91, edt.\ by R.K. Ellis et
al.\ (World Scientific, 1991)}
Grinstein,\footnote{in: {\sl High Energy Phenomenology}, edt. by R.
Huerta and M.A. P\'eres (World Scientific, 1991)}
Isgur and Wise,\footnote{in: {\sl Heavy Flavours}, edt.\ by A.J. Buras
and M. Lindner (World Scientific, 1992)}
and Mannel.\footnote{to appear in: {\sl QCD--20 Years Later}, edt.\ by
P.M. Zerwas and H.A. Kastrup (World Scientific)}
They were very helpful in preparing the material for these chapters.
The second part (Chapters~3 and 4) is devoted to a very detailed and
comprehensive description of the formalism of heavy quark effective
theory, and of the state of the art in the analysis of symmetry
breaking corrections. The discussion of these topics is necessarily
more advanced and, with the exception of the introductory sections
3.1--3.4 and 4.1--4.3, addresses the experts in the field. Some of the
results presented in these chapters have not even been published so
far. In particular, we present for the first time the complete
expressions for meson and baryon weak decay amplitudes to order $1/m_Q$
in the heavy quark expansion, and to next-to-leading order in QCD
perturbation theory. The third part consists of Chapter~5, in which we
discuss nonperturbative techniques which aim at a dynamical, QCD-based
calculation of hadronic matrix elements. We give a general introduction
to the QCD sum rule approach of Shifman, Vainshtein, and Zakharov,
which might help readers not familiar with this subject to get an
appreciation of the physical motivations of the method. In the
subsequent sections we present many interesting results that have been
obtained recently by applying the sum rule technique to the heavy quark
effective theory. In the last part (Chapter~6) we come back to the
phenomenological applications of heavy quark symmetry. We present a
comprehensive analysis of semileptonic $B$ meson decays in the context
of the new theoretical framework provided by the heavy quark effective
theory. The virtue of this approach is that model independent aspects
of the analysis can be clearly separated from model dependent ones.
Ultimately, this might help to change the perspective in heavy quark
phenomenology from a comparison of the data with models to a language
which deals with certain types of corrections in a well-defined
expansion in QCD. This would be a significant development.

\end{document}